\documentstyle[prl,epsf,amssymb,amsbsy,amstext,amsfonts,aps]{revtex}
\begin{document}
\twocolumn
\draft{}

\bibliographystyle{try}

\topmargin 0.0001cm

 
 \newcounter{univ_counter}
 \setcounter{univ_counter} {0}

\addtocounter{univ_counter} {1} 
\edef\INFNGE{$^{\arabic{univ_counter}}$ } 

\addtocounter{univ_counter} {1} 
\edef\SACLAY{$^{\arabic{univ_counter}}$ } 

\addtocounter{univ_counter} {1} 
\edef\MSU{$^{\arabic{univ_counter}}$ }

\addtocounter{univ_counter} {1} 
\edef\ASU{$^{\arabic{univ_counter}}$ } 

\addtocounter{univ_counter} {1} 
\edef\CMU{$^{\arabic{univ_counter}}$ } 

\addtocounter{univ_counter} {1} 
\edef\CUA{$^{\arabic{univ_counter}}$ } 

\addtocounter{univ_counter} {1} 
\edef\CNU{$^{\arabic{univ_counter}}$ } 

\addtocounter{univ_counter} {1} 
\edef\UCONN{$^{\arabic{univ_counter}}$ } 

\addtocounter{univ_counter} {1} 
\edef\DUKE{$^{\arabic{univ_counter}}$ } 

\addtocounter{univ_counter} {1} 
\edef\FIU{$^{\arabic{univ_counter}}$ } 

\addtocounter{univ_counter} {1} 
\edef\FSU{$^{\arabic{univ_counter}}$ } 

\addtocounter{univ_counter} {1} 
\edef\GWU{$^{\arabic{univ_counter}}$ } 

\addtocounter{univ_counter} {1} 
\edef\ORSAY{$^{\arabic{univ_counter}}$ } 

\addtocounter{univ_counter} {1} 
\edef\ITEP{$^{\arabic{univ_counter}}$ } 

\addtocounter{univ_counter} {1} 
\edef\INFNFR{$^{\arabic{univ_counter}}$ }

\addtocounter{univ_counter} {1} 
\edef\JMU{$^{\arabic{univ_counter}}$ } 

\addtocounter{univ_counter} {1} 
\edef\KYUNGPOOK{$^{\arabic{univ_counter}}$ } 

\addtocounter{univ_counter} {1} 
\edef\MIT{$^{\arabic{univ_counter}}$ } 

\addtocounter{univ_counter} {1} 
\edef\UE{$^{\arabic{univ_counter}}$ } 

\addtocounter{univ_counter} {1} 
\edef\UMASS{$^{\arabic{univ_counter}}$ }

\addtocounter{univ_counter} {1} 
\edef\UNH{$^{\arabic{univ_counter}}$ } 

\addtocounter{univ_counter} {1} 
\edef\NSU{$^{\arabic{univ_counter}}$ } 

\addtocounter{univ_counter} {1} 
\edef\OHIOU{$^{\arabic{univ_counter}}$ } 

\addtocounter{univ_counter} {1} 
\edef\ODU{$^{\arabic{univ_counter}}$ } 

\addtocounter{univ_counter} {1} 
\edef\PITT{$^{\arabic{univ_counter}}$ } 

\addtocounter{univ_counter} {1} 
\edef\RPI{$^{\arabic{univ_counter}}$ } 
 
\addtocounter{univ_counter} {1} 
\edef\RICE{$^{\arabic{univ_counter}}$ } 

\addtocounter{univ_counter} {1} 
\edef\URICH{$^{\arabic{univ_counter}}$ } 

\addtocounter{univ_counter} {1} 
\edef\SCAROLINA{$^{\arabic{univ_counter}}$ } 

\addtocounter{univ_counter} {1} 
\edef\UTEP{$^{\arabic{univ_counter}}$ } 

\addtocounter{univ_counter} {1} 
\edef\JLAB{$^{\arabic{univ_counter}}$ } 

\addtocounter{univ_counter} {1} 
\edef\VT{$^{\arabic{univ_counter}}$ } 

\addtocounter{univ_counter} {1} 
\edef\VIRGINIA{$^{\arabic{univ_counter}}$ } 

\addtocounter{univ_counter} {1} 
\edef\WM{$^{\arabic{univ_counter}}$ } 

\addtocounter{univ_counter} {1} 
\edef\YEREVAN{$^{\arabic{univ_counter}}$ }

\title{Photoproduction of the $\omega$ meson  on the proton at large momentum transfer}

 \author{ 
M.~Battaglieri,\INFNGE\
M.~Brunoldi,\INFNGE\
R.~De Vita,\INFNGE\
J.M.~Laget,\SACLAY\
M.~Osipenko,\INFNGE\
M.~Ripani,\INFNGE\
M.~Taiuti,\INFNGE\
G.~Adams,\RPI\
M.J.~Amaryan,\YEREVAN\
E.~Anciant,\SACLAY
% ...
\
\\(CLAS Collaboration)
}

\address{\INFNGE Istituto Nazionale di Fisica Nucleare, Sezione di Genova and Dipartimento di Fisica, Universit\`a di Genova, Italy 16146}
\address{\SACLAY CEA-Saclay, Service de Physique Nucleaire, Gif-sur-Yvette, France 91191}
\address{\MSU Moscow State University,Moscow, Russia 119899}
\address{\ASU Arizona State University, Tempe, Arizona 85287-1504}
\address{\CMU Carnegie Mellon University, Pittsburgh, Pennsylvania 15213}
\address{\CUA Catholic University of America, Washington, D.C. 20064}
\address{\CNU Christopher Newport University, Newport News, Virginia 23606}
\address{\UCONN University of Connecticut, Storrs, Connecticut 06269}
\address{\DUKE Duke University, Durham, North Carolina 27708-0305}
\address{\FIU Florida International University, Miami, Florida 33199}
\address{\FSU Florida State University, Tallahasee, Florida 32306}
\address{\GWU The George Washington University, Washington, DC 20052}
\address{\ORSAY Institut de Physique Nucleaire d'Orsay, IN2P3, BP 1, Orsay, France 91406}
\address{\ITEP Institute of Theoretical and Experimental Physics, Moscow, Russia, 117259}
\address{\INFNFR Istituto  Nazionale di Fisica Nucleare, Laboratori Nazionali di Frascati, Frascati, Italy 00044}
\address{\JMU James Madison University, Harrisonburg, Virginia 22807}
\address{\KYUNGPOOK Kyungpook National University, Taegu 702-701, South Korea}
\address{\MIT Massachusetts Institute of Technology, Cambridge, Massachusetts  02139-4307}
\address{\UE University of Edinburgh, Edinburgh, Scotland, UK EH9 3JZ}
\address{\UMASS University of Massachusetts, Amherst, Massachusetts  01003}
\address{\UNH University of New Hampshire, Durham, New Hampshire 03824-3568}
\address{\NSU Norfolk State University, Norfolk, Virginia 23504}
\address{\OHIOU Ohio University, Athens, Ohio  45701}
\address{\ODU Old Dominion University, Norfolk, Virginia 23529}
\address{\PITT University of Pittsburgh, Pittsburgh, Pennsylvania 15260}
\address{\RPI Rensselaer Polytechnic Institute, Troy, New York 12180-3590}
\address{\RICE Rice University, Houston, Texas 77005-1892}
\address{\URICH University of Richmond, Richmond, Virginia 23173}
\address{\SCAROLINA University of South Carolina, Columbia, South Carolina 29208}
\address{\UTEP University of Texas at El Paso, El Paso, Texas 79968}
\address{\JLAB Thomas Jefferson National Accelerator Facility, Newport News, Virginia 23606}
\address{\VT Virginia Polytechnic Institute and State University, Blacksburg, Virginia   24061-0435}
\address{\VIRGINIA University of Virginia, Charlottesville, Virginia 22901}
\address{\WM College of Willliam and Mary, Williamsburg, Virginia 23187-8795}
\address{\YEREVAN Yerevan Physics Institute, Yerevan, Armenia 375036 }

\date{\today}

\maketitle

\newpage

\wideabs{
\begin{abstract}
The differential cross section, $d\sigma/dt$ for $\omega$ 
meson exclusive photoproduction on the proton above the resonance 
region ($2.6<W<2.9$ GeV) was  measured
up to a momentum transfer $-t = 5$ GeV$^2$ 
using the CLAS detector
at  Jefferson Laboratory.
The $\omega$ channel was identified by detecting a
proton and  $\pi^+$ in the final state and 
using the missing mass technique.
While the low momentum transfer region shows 
the typical diffractive pattern expected from Pomeron and
Reggeon exchange, at large $-t$
the differential cross section has a flat behavior. This
feature can be explained by introducing 
quark interchange processes in addition to the QCD-inspired two-gluon exchange.

\end{abstract}

\pacs{PACS : 13.60.Le , 12.40.Nn, 13.40.Gp}
}

\narrowtext

\begin{figure}[h]
\epsfxsize9.5cm
\epsfysize6.3cm
\centerline{\epsffile{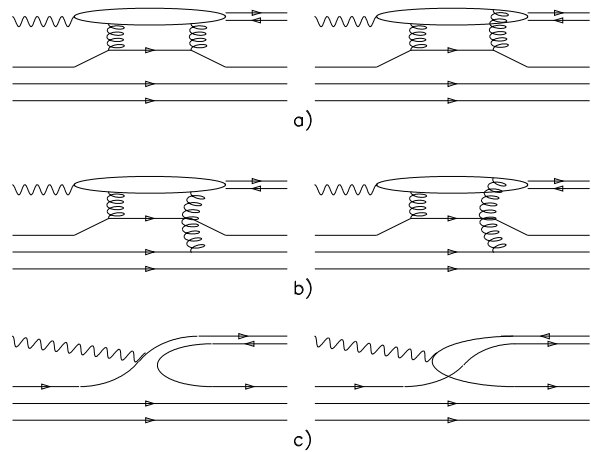}}
\caption[]{
The Feynman diagrams corresponding to 
a) two-gluon exchange from a single quark,
b) two-gluon exchange taking into account quark correlations in the nucleon,
and c) quark exchange.
}
\label{diagrams}
\end{figure}

In this paper we report results of the first comprehensive measurement  
of the cross section for  $\omega$ meson photoproduction 
on protons  for  $E_\gamma$  between 3.19 and 3.91 GeV 
over the  $-t$ range $0.1-5.0$ GeV$^2$. 
Previous studies at DESY, SLAC, and NINA~\cite{Abbhhm,Ba73,Cl77}
are sparse and cover a limited kinematic range of
$-t < 1$ GeV$^2$ ~\cite{Abbhhm,Ba73} and $-t\sim-t_{max}$ (4-5 GeV$^2$)  ~\cite{Cl77}.
The low momentum transfer data  ($-t < 1$ GeV$^2$) shows a diffractive behavior
that can be 
interpreted in the framework of the Vector Meson Dominance (VMD) model~\cite{Ba78}
as the elastic scattering of  vector mesons off the proton target.
In a more recent approach,  this process is also described by the 
$t$-channel exchange of the Pomeron 
and the dominating $\pi$ Regge trajectory~\cite{La00}.
Other approaches~\cite{Fr96,Oh} based on effective Lagrangians
and inclusion of nucleon resonances as predicted by quark model calculations,
are able to  reproduce the data at lower photon energies. 
At high $-t$, where the cross section  is sensitive to the microscopic details 
of the interaction, the underlying physics can be described  using 
parton degrees of freedom. The onset of this regime can be tested by a 
combined analysis of different flavor channels.
The recent JLab  measurements of  $\phi$~\cite{An00}
and  $\rho$~\cite{Ba01}  photoproduction cross sections at large momentum
transfer show a behavior consistent with 
a  QCD-inspired framework~\cite{DL87,La95,Ca02}. At large $-t$,  the small impact
parameter ($\approx 1/{\sqrt{-t}}$)  prevents the constituent gluons (quarks) of the exchange
from interacting and forming a Pomeron (Reggeon).  
Because of the dominant $s\overline{s}$ component of the 
$\phi$,  quark exchange is strongly suppressed in this channel by the OZI rule  
and the two-gluon mechanism dominates (Fig.~\ref{diagrams}-a-b)~\cite{La00,La95,La98}.
In contrast, the light quark composition of the $\rho$  allows  valence
quarks to be exchanged between the baryon  and the meson states
(Fig.~\ref{diagrams}-c)~\cite{La00,Ca02}. The same quark exchange mechanism is predicted
to dominate the $\omega$ sector.
Complete and detailed measurements of the $\omega$ differential cross section
are therefore  a stringent test of this conjecture.

The measurement was performed at Jefferson Lab with a  bremsstrahlung 
photon beam  produced by a continuous electron beam of  $E_0$ = 4.1 GeV  hitting 
a gold foil of   $10^{-4}$ radiation lengths. 
A bremsstrahlung  tagging system~\cite{SO99},
 with a photon energy resolution of 0.1$\%$ $E_0$, 
was used to tag photons in the energy range from $3-4$ GeV.
The target cell, a mylar cylinder of  6 cm in diameter and 18 cm long, was filled
with liquid hydrogen at 20.4 K.
The high-intensity 
photon flux ($\sim 4\cdot 10^6 \gamma$/s) was continuously monitored during  data taking by an $e^+ e^-$  pair spectrometer 
located downstream of the target. 
The efficiency of this device was determined during dedicated low intensity ($\sim 10^5 \gamma$/s )
runs  by comparison with  a 100\% efficient lead-glass total absorption counter.
The systematic uncertainty of the photon flux has been estimated to be 5$\%$.

The hadrons were detected in CLAS~\cite{B00} (CEBAF Large Acceptance Spectrometer), a spectrometer with nearly  
4$\pi$ coverage  with a toroidal magnetic field ($\sim 1 T$) generated
by six superconducting coils. 
The field was set to bend the positive particles
away from the beam into the acceptance of the 
detector.
Three drift chamber regions allowed tracking of charged particles~\cite{DC}, and
time-of-flight scintillators (TOF) were used for hadron
identification~\cite{Sm99}. The momentum resolution  was of the order of a few percent,
while the detector geometric acceptance  
was about 70\% 
for positive hadrons.
Low energy negative particles, however, were mainly lost at 
forward angles because they were bent out of the acceptance.
Coincidences between the photon tagger and the CLAS detector (TOFs)
triggered the recording of hadronic interactions. From a total of  70M triggers,
100k events were identified as $p \omega$ candidates.
\begin{figure}[h]
\epsfxsize8.cm
\epsfysize7.7cm
\centerline{\epsffile{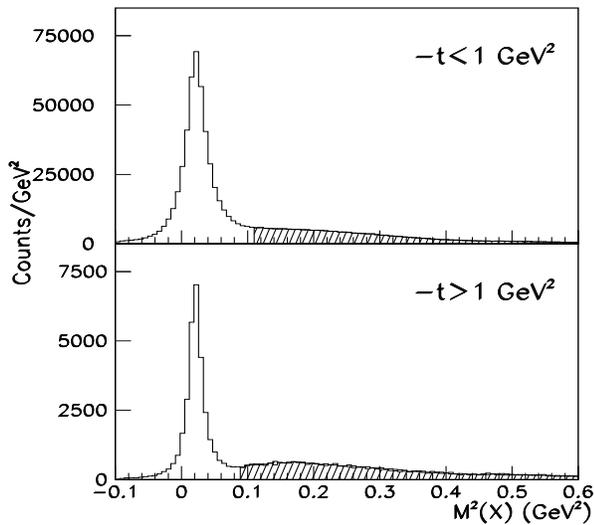}}
\caption[]{Missing mass squared for the reaction $\gamma p \rightarrow  p \pi^+ X$ with  
$E_\gamma$  between 3.19 and 3.91 GeV. The hatched area corresponds to the $\omega$ candidates.}
\label{miss2_p_pi}
\end{figure}

For this analysis we chose the 
most sizeable $\omega$ decay mode
($\omega \rightarrow \pi^+  \pi^- \pi^0$, b.r. 88.8$\%$),
requiring  detection of both the proton and the $\pi^+$ in CLAS.
The data analysis consisted of two main steps:
two-pion background rejection and
$\omega$ yield extraction from the multi-meson background.
Due to the different dynamics governing the low and the high $-t$
 domains, we divided the data set into two samples corresponding
to low ($-t<1$ GeV$^2$) and high ($-t>1$ GeV$^2$) momentum transfer.
The analysis procedure was then performed and optimized 
independently for the two samples.

The two-pion background is dominated by the $\gamma p \rightarrow  p \rho^0$
channel since its cross section is five times larger than that for $\gamma p \rightarrow  p \omega$
for $E_{\gamma}\sim$3-4 GeV, and 
the mass of the $\rho$ meson (770 MeV) is very close to the 
$\omega$ mass (783 MeV). Even though  the $\rho$ has a larger width ($\sim$150 MeV FWHM)
compared to the $\omega$ ($\sim$8 MeV FWHM enlarged to $\sim$55 MeV FWHM by the experimental 
resolution), the missing mass for the reaction $\gamma p \rightarrow  p X$  alone
does not allow  separation of the two channels. 
The two-pion background was rejected by requiring that the  missing
mass for the reaction $\gamma p \rightarrow  p \pi^+ X$ be larger than 0.3 GeV.
We estimated that the $\omega$'s surviving this cut were around 99$\%$.
Figure~\ref{miss2_p_pi} shows the ($p \pi^+$) 
missing mass squared spectrum:  the missing $\pi^-$ peak was easily removed 
(the hatched area corresponds to the retained events).
The small
contamination surviving the cut 
(estimated to be around 5$\%$ by the 
simulations) is  spread over  a wide proton missing mass interval, and it was
reduced to a  negligible level  in the second step of the analysis.
The  $\omega$ yield extraction from the multi-meson background 
was performed on the proton missing mass spectrum
by using two different procedures:
a Gaussian fit to the $\omega$ peak and
a side-band subtraction.

\begin{figure}[h]
\epsfxsize8.cm
\epsfysize7.7cm
\centerline{\epsffile{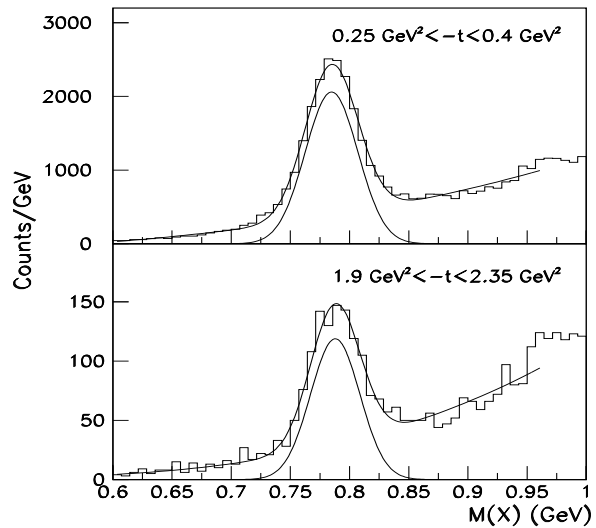}}
\caption[]{Missing mass for $\gamma p \rightarrow  p + X$ around the $\omega$ mass with $E_\gamma$  between 3.74 
and 3.92 GeV. The spectra are fitted to a Gaussian + 4$^{th}$-order polynomial.}
\label{miss_p_sb_g}
\end{figure}

Both of them rely on the hypothesis of a smooth and continuous
background variation from one side-band region to the other.
The two methods were not totally  independent, 
but the comparison of their 
results allowed estimation of the systematic error related 
to the $\omega$ identification.
The proton missing mass in each $-t$ bin was fitted  to a Gaussian curve  
(the $\omega$ peak)  plus a fourth order polynomial (the multi-meson background).
The $\omega$ yields in each $-t$ bin were the area under the  Gaussian.
Figure~\ref{miss_p_sb_g} shows the fitted spectra in a low and high  $-t$ bin.
The side-band subtraction procedure allowed  extraction of  a localized signal over an extended
background  subtracting the regions  on either side of the peak (side bands)
after a proper normalization. 
The middle region was fixed at 6$\sigma$ of the Gaussian curve describing
the $\omega$ peak ($\sigma\sim24$ MeV) while the side bands had a width of 3$\sigma$ each.
The $\omega$ yield was obtained as average of the two procedures while the
maximum difference, 8$\%$, was used as an estimate of the systematic error.

The CLAS acceptance and reconstruction efficiency  were evaluated with  Monte Carlo simulations
using the event generator of Ref.~\cite{Co94}. This code included  the  main contributions to 
the $p \pi^+  \pi^-$  ($\gamma p \rightarrow  p \rho^0$, $\gamma p \rightarrow   \Delta^{++} \pi^-$,
 and $\gamma p \rightarrow  p \pi^+  \pi^-$ in $s$-wave) and  $p \pi^+  \pi^-  \pi^0$ final states
($\gamma p \rightarrow  p \omega$ and $\gamma p \rightarrow  p \pi^+  \pi^- \pi^0$ phase space), 
along with background reactions with 
four  or more pions.
The generated events were processed by a GEANT-based code simulating the CLAS detector,
and reconstructed using the same analysis procedure that was applied to the raw data.
The acceptance was derived as a
function of $E_\gamma$ and the momentum transfer $t$, integrating over the remaining 
independent variables. 
To minimize the model dependence in the
acceptance calculation, the $\gamma p \to p \omega$ differential 
cross section was iteratively determined from the data and implemented
in the Monte Carlo code.
The final state  ($p$ and $\pi^+$ detected)
did not allow us to measure the $\omega$ decay, therefore the available experimental data about the 
decay matrix elements~\cite{Abbhhm,Ba73}, as well as the general decay property
of vector mesons~\cite{St61}, was implemented in the event generator.
The systematic error associated  with the efficiency calculation was estimated by
comparing the results  obtained after generating events with slightly different distributions
both in production and decay. 
The resulting systematic uncertainty  was  estimated to be $\sim 10\%$.
The average acceptance of CLAS for detected $p \pi^+$ ranged from $8\%$ to $10\%$.
For the very forward angles ($-t<0.1$ GeV$^2$) and the very backward angles
($-t\sim-t_{max}$) the CLAS detector had no acceptance  for
this reaction. 

\begin{figure}[h]
\epsfxsize9.cm
\epsfysize10.cm
\centerline{\epsffile{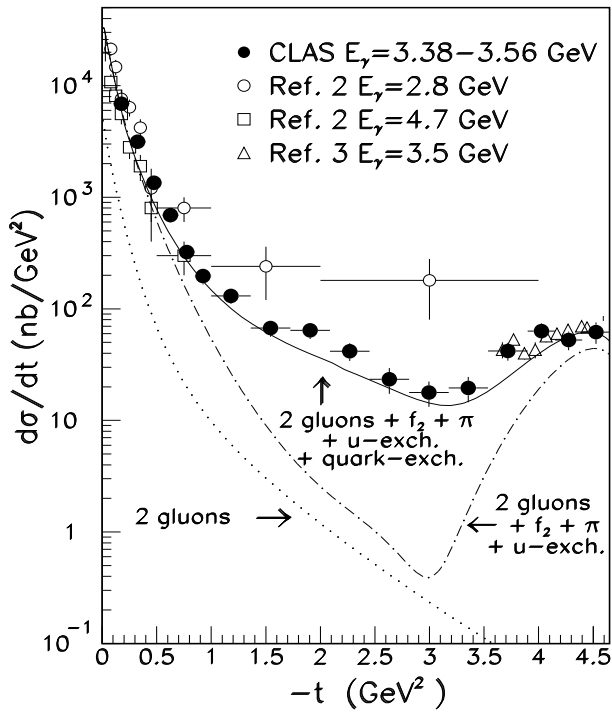}}
\caption[]{Differential cross section for $\gamma p \to p \omega$ 
as measured in CLAS for the energy bin $E_{\gamma}$=3.38-3.56~GeV 
 compared with existing data. See the text for the explanation of the curves. 
In this energy bin, $\theta_{\omega}^*$=90$^{\circ}$ corresponds to $-t$=2.52 GeV$^2$.}
\label{ds_omg}
\end{figure}
\begin{figure}[h]
\epsfxsize9.cm
\epsfysize10.cm
\centerline{\epsffile{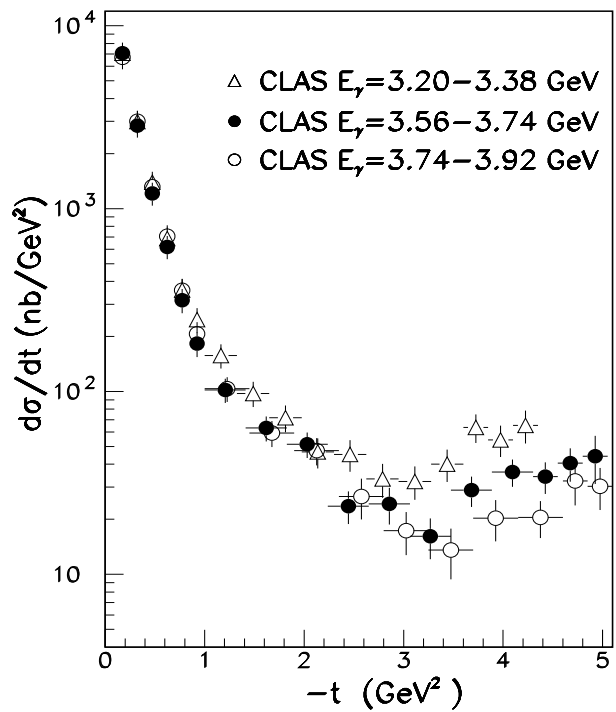}}
\caption[]{Differential cross section measured in CLAS. 
The fourth energy bin ($E_\gamma=3.38-3.56$ GeV) is shown in Fig.~\ref{ds_omg}.}
\label{omg_w_dep}
\end{figure}

The $\omega$ photoproduction cross section as a function of $t$ 
was extracted in four energy bins in the range 3.19-3.91 GeV.
Data are shown in  Fig.~\ref{ds_omg} and~\ref{omg_w_dep}:
vertical error bars  include both the statistical 
uncertainties (ranging from 2$\%$ to 25$\%$) and the overall systematic
error (14$\%$) summed in quadrature, while the horizontal bars 
reflect the bin sizes. 
In the low momentum transfer region, $0.1<-t<0.5$ GeV$^2$, good agreement with 
the previous measurement  of Ref.~\cite{Ba73} in a similar energy range is evident.
At higher $-t$  the CLAS data lie between the two data sets taken respectively
at smaller and larger energy. 
Assuming an exponential $A e^{Bt}$  behavior in the range  $0.1<-t<0.5$ GeV$^2$, 
the  coefficient resulting from this experiment,
$B =5.4\pm0.6$ GeV$^{-2}$,  is consistent with the values  $B=5.1\pm1.4$ GeV$^{-2}$
and  $B =7.1\pm1.7$ GeV$^{-2}$ obtained by fitting respectively, the $E_\gamma=2.8$ GeV
and $E_\gamma=4.7$ GeV data sets reported in Ref.~\cite{Ba73}. 
Good agreement is also found with  existing  data at the largest momentum transfer  
taken at NINA~\cite{Cl77}  
with a bremsstrahlung photon beam  and  a single arm spectrometer.

Predictions of the  QCD-inspired model of  Refs.~\cite{La00,Ca02} 
are also  shown in Fig.~\ref{ds_omg}.
Here the Pomeron exchange has been  replaced 
by the exchange of two non-perturbatively dressed gluons (dotted line).
The low momentum transfer region is dominated by the pion exchange
that, added to the two-gluon and $f_2(1270)$ trajectory exchanges,
gives  good agreement up to $-t\sim$ 1 GeV$^2$.
The $\pi$ exchange   gives a strong contribution because of the large
coupling constant $g_{\omega\pi\gamma}$ (0.334).
Close to the upper kinematic limit ($-t\sim-t_{max}$) the cross section
is well reproduced by the exchange  of the nucleon
Regge trajectory in the $u$ channel~\cite{Gu97}. 
At intermediate momentum transfer, the two gluon exchange contribution
underestimates (by an order of magnitude) the experimental cross section.
The calculation uses the same expression as in  our phi-photoproduction 
work~\cite{La00,An00}, where only the relevant mass and radiative decay 
width have been changed. 
In contrast to the $\phi$ meson, quark interchanges (Fig.~\ref{diagrams}-c)
are not forbidden in  $\omega$ production.
As explained in Refs.~\cite{Gu97,Co84} these hard-scattering mechanisms can be
incorporated in an effective  way by using the so called ``saturated'' trajectory
that is independent of $t$ at large momentum transfer~\cite{Se94}.
Regge trajectories are usually assumed to be linear in $t$, but there are both phenomenological
and theoretical arguments supporting the idea of non-linear trajectories~\cite{Br00}.
Saturated trajectories lead to the asymptotic quark counting rules~\cite{Br73}
that, model independently, determine the energy behavior of the 
cross section at large $-t$.
This approach was successfully adopted to explain the large momentum transfer 
$hadron-hadron$ interactions~\cite{Co78,Fi99,Wh94}, as well as 
several photon-induced reactions~\cite{Ba01,Gu97,Sh79}. 
The pion saturating trajectory 
(\mbox{$\alpha_{\pi}^{sat}(t)\rightarrow$ -1} when $-t\rightarrow$ - $\infty$)
is in a  form that reproduces
 the $\gamma p \rightarrow n \pi^+$ 
reaction around $\theta_{\omega}^*$=90$^{\circ}$~\cite{Gu97}.
The solid line in Fig.~\ref{ds_omg} shows the full calculation,
including such a  saturating trajectory, while the dot-dashed line corresponds
to the same calculation with linear trajectories. 
Quark exchange  increases the cross section
at large $-t$ by more than one order of magnitude.

The  measured   $d\sigma/dt$ in the other three
photon energy bins are shown in Fig.~\ref{omg_w_dep}.
From the four data sets, the  cross section at
$\theta_{\omega}^*$=90$^{\circ}$  was extracted as a function of energy.
A power law fit $s^{-C}$ to $d\sigma/dt$ at $\theta_{\omega}^*$=90$^{\circ}$ was
performed also using the only other datum available in the literature 
(SLAC datum at $s$ = 6.13 GeV$^2$~\cite{Ba73}).
The experimental points include both statistical and systematic errors
summed in quadrature.
The fit yields $C=7.2\pm0.7$ ($\chi^2=0.5$).
It is the  first time that such a power law
behavior, seen for other exclusive reactions~\cite{Ba73,Sh79},
has been observed in the $\omega$ channel.
The quark exchange diagrams 
of Fig.~\ref{diagrams}-c-left (point-like interaction) and~\ref{diagrams}-c-right
(hadronic component of the photon)  have
a $s^{-7}$ and $s^{-8}$  power-law behavior, respectively, both in dimensional counting~\cite{Br73} and 
in recent models~\cite{Ha00}.
Note that  the saturated $\pi$ Regge trajectory behaves like $s^{-8}$, too.
Besides the differential cross section at fixed energy, 
the $s$ dependence is a strong hint of  the presence of  quark interchange
hard mechanisms in addition to the  two-gluon exchange process.

In conclusion, elastic photoproduction of the $\omega$ mesons from the proton
was measured for the first time with nearly complete kinematic coverage. 
The energy power law behavior of the differential cross section at $\theta_{\omega}^*$=90$^{\circ}$
 was observed.
The comparison  with a QCD-inspired model, able to reproduce the  $\phi$ and the $\rho^0$
photoproduction data, provides  further evidence  for the presence of hard processes.
Adopting  a QCD language in this energy region, the two-gluon exchange mechanism
(that fully  describe the $\phi$ photoproduction data) badly miss  
the cross section at large momentum transfer
and its energy dependence.
Good agreement is achieved when 
quark interchange processes, suppressed in the $\phi$ channel and  weakly contributing in the $\rho$ case, 
are  included in an effective way  in the calculation of the $\omega$ cross section.

We would like to acknowledge the outstanding efforts of the staff of the Accelerator
and the Physics Divisions at JLab that made this experiment possible. 
This work was supported in part by  the  Italian Istituto Nazionale di Fisica Nucleare, 
the French Centre National de la Recherche Scientifique and the Commissariat \`{a} l'Energie Atomique, 
the U.S. Department of Energy and the National Science Foundation, 
and the Korea Science and Engineering Foundation.
The Southeastern Universities Research Association (SURA) operates the
Thomas Jefferson National Accelerator Facility for the United States
Department of Energy under contract DE-AC05-84ER40150.

\end{document}